# Electrical effects in superfluid helium.
# I. Thermoelectric effect in Einstein's capacitor.

Dimitri O. Ledenyov, Viktor O. Ledenyov, Oleg P. Ledenyov

*National Scientific Centre Kharkov Institute of Physics and Technology,
Academicheskaya 1, Kharkov 61108, Ukraine.*

The *Einstein's* research ideas on the thermodynamical fluctuational nature of certain electrical phenomena [1] and the physical nature of the electric potentials difference $U$ in an electric capacitor at the temperature of $T$ [2] were proposed in *1906-1907*. On the base of *Einstein's* research ideas, we explain the recent experimental results [3, 4] and propose the consistent theory on the physical nature of the electric effects in an electric capacitor at an action of the second sound standing wave in the *superfluid Helium ($^4He$)* and in the rotational torsional mechanical resonator in the *Helium II*. The use of the *Einstein's* research approach, based on the consideration of an interconnection between the thermal, mechanical and electrical fluctuations, allows us to obtain the quantitative theoretical research results, which are in a good agreement with the experimental data on the correlations of the alternate low temperatures difference in the second sound wave, and the alternate electric potentials difference between the capacitor plates in the *superfluid Helium* as well as in the rotational torsional mechanical resonator in the *Helium II*.

## Introduction

In the research paper on the *Brownian* motion [1] in *1906*, *Einstein* first showed that there is an interconnection between the thermodynamical fluctuations and the electric effects in the linear electric circuit. In an agreement with the *Einstein's* research conclusions, the electric charge $q(t)$, passing through the cross section of closed circuit conductor, is defined by the temperature $T$, the closed circuit resistance $R$ and the observation time $t$, in the expression $\overline{q^2} = 2t\, k_B T / R$, where $k_B$ is the *Boltzmann* constant, the upper dash designates the averaged value. In this electric circuit, the expression for the current is $\overline{I^2} = 2k_B T / Rt$, and the equation for the difference between the electric potentials is $\overline{U^2} = 2R\, k_B T / t$. In *1907*, *Einstein* [2] demonstrated that the electric potentials difference $U$ at the plates of an electric capacitor with the capacitance $C$ must appear, because of the thermodynamical fluctuational effects in the surrounding physical medium at the temperature $T$. *Einstein* [2] proposed that the electric capacitor can be considered as a *Brownian* particle, which has the one degree of freedom, connected with its electric subsystem; and in an agreement with the *Einstein's* calculations, the fluctuations of the electric potentials difference $U$ can be observed in an electric capacitor. According to the equipartition theorem for the system with any degree of freedom, it appears that the electric energy, stored by an electric capacitor, is equal to the thermal energy as written in the equation $C\overline{U^2}/2 = k_B T/2$.

In [2], the corresponding calculations were completed, and it was shown that the electric potentials difference $U$ can be measured with the use of experimental measurement setup at appropriate selection of the capacitance $C$ at the room temperature $T$. On that time, the discussed experiment [2] was not realized, because the *Einstein's* first research paper on the thermodynamical electro-physics [2] was not appropriately noticed among a number of his other revolutionary research ideas. A few years later, *de Haas-Lorentz* [5] developed the similar theoretical representations for other elements of linear electric circuits and showed that, in the fluctuational case, the inductance $L$ will store the energy as described in the equation $L\overline{I^2}/2 = k_B T/2$. In all the considered cases, the thermodynamic fluctuations can induce the electrical currents as well as dissipate the electrical currents in the electric circuits. The heating energy, transforming at this process over the time $\tau$, is equal to $\overline{U^2}\tau/R = 4k_B T$ (in this case, the electric potentials difference $U = IR$, $I$ is the electric current, $\tau$ is the time, $\tau << R/L$). The deviation of the order of magnitude of obtained result in the two times in the given expression in comparison with the data in the expression [1] is connected with the fact that the values of the characteristic time periods for the linear magnitude of the electric potentials difference $U$ and the quadratic magnitude of the electric potentials difference $U^2$ differ in the two times. In *1928*, *Nyquist* [6] considered the spectrum of possible frequencies of electrical oscillations in the linear *RLC* electric circuit, and found the expression for the existing interconnection between the time $\tau$ and the frequencies band span in which the registration of signal is per-formed $\Delta\omega=2\pi/\tau$. *Nyquist* [6] also pointed to the possible quantum case, which can be realized at the following condition $\hbar\omega>k_B T$, where $\omega$ is the circular frequency of the electric signal in a linear circuit. In *1951*, *Callen* and *Welton* [7] obtained the quantum fluctuation – dissipation formula with the consideration of contribution of zero oscillations, and found the important expression $\overline{U_\omega^2}=\hbar\omega R_\omega cth(\hbar\omega/2k_B T)$ (also see [8-11]), where $\hbar=h/2\pi$, $h$ is the *Planck* constant. As far as we know, the experiment for linear electric circuit with a capacitor, proposed by *Einstein* [2], was not completed in the *XX* century. At the same time, the fluctuational electric noises in the resistors were researched widely. The results obtained in the fluctuation – dissipation theory for the linear electric circuits were confirmed in both the classic case as well as the quantum case, serving as one of a number of possible methods to research the electromagnetic fluctuations and the *Van der Waal's* forces [12, 13].

In *2004*, *Rybalko* [3] measured the alternating difference of electric potentials $U'$ between the plates of an electric capacitor in the superfluid *Helium II* at the temperature of $T<T_\lambda$ ($T_\lambda=2.172K$ is the temperature of superfluid helium transition). The wave of second sound with the frequency of $\omega_2$ was propagating in *Helium II*, and was accompanied by the oscillations of the temperatures difference $T'$. The amplitude of oscillations of temperature difference was equal to $T'\approx(10^{-2}-10^{-4})K$, and, as it was discovered in [3], the alternating difference of electric potentials with the magnitude of $U'\approx(10^{-7}-10^{-9})V$ was generated at the same frequency of $\omega_2$. The relation of amplitudes satisfied the condition $U'/T'\approx k_B/2e$, where $e$ is the electrical charge of an electron.

In *2005*, *Rybalko, Rubets* [4] conducted an experiment with the electric capacitor placed into the torsional mechanical resonator with *Helium II*, which oscillated with the frequency of $\Omega$. In this case, the alternating electric potentials difference $U'$ with the approximately same amplitude at the frequency of $2\Omega$ was registered. Authors [3, 4] proposed that the origination of the electric potentials difference $U'$ is due to some effects, connected with the electric polarization of *Helium II*, however a possible physical mechanism of electric polarization of *Helium II* is not known yet.

Presently, there are more than the ten published theoretical researches [14–26], which consider the possible scientific interpretation of the experimental data in [3, 4]. In these research papers, the propositions about the possibility of existence of some new effects, including the inertial effects [16], which might lead to the electric activity and polarization of superfluid helium in conditions of low temperature experiments in [3, 4], are made. However, the simple mathematical analysis [21, 22] on the proposed theoretical models demonstrates that the magnitude of oscillations of electric potentials difference $U'$ is in $10^5-10^6$ times less in the proposed theoretical models [14–26] than the magnitude of the amplitude, which was experimentally observed in [3, 4]. Also, it is not explained: why does the magnitude of oscillations of electric potentials difference $U'$ not depend on the temperature $T$ in the experiment in [3], when all the properties of *Helium II* must change in the researched range of temperatures strongly?

We will show up that some of the experimentally observed effects [3, 4], in our opinion, are connected with the thermoelectric fluctuational effect, appearing in the electric capacitor, when there is an alternating difference of temperatures $T'$ between the plates of an electric capacitor, first considered by *Einstein* [2]. Let us take into the attention the fact that the conditions $(\hbar\omega_2, \hbar\Omega)<<k_B T$ are true in the case of the second sound wave and rotational oscillations in [3, 4], therefore these effects belong to the classic case of fluctuational processes, hence it is enough to use the representations by *Einstein* in [2] to make their detailed theoretical description.

## Fluctuational thermoelectric effect in electric capacitor

As it is well known, any physical body with the temperature of $T$, is a source of the electromagnetic radiation. The spectrum of electromagnetic radiation depends on the physical body's temperature only, if the electromagnetic body and radiation are in the state of thermal equilibrium. Following the *Einstein's* research ideas in [2], let us consider the case, when the characteristic radiation frequency $\omega$ satisfies the classic condition $\hbar\omega<k_B T$. We will believe that the quasi-statistical electric and magnetic fields are dependent on the thermal fluctuations of electric charges. The state of physical system is characterized by some magnitudes of its physical parameters $\xi_1, \xi_2, ...\xi_n$, defining the thermodynamic state [2, 27]. In a general case, these parameters can characterize the microscopic state of physical body as well as the macroscopic state of physical body. In the microscopic case, when the physical body can be represented by a group of microscopic particles $n$, the value of physical parameter $\xi_i$ will characterize the particle $i$. In the macroscopic case, the value of physical parameter $\xi_i$ will characterize all the physical body with the different states of freedom with the indexes of $i$. In the equilibrium state, these parameters have some average values $\overline{\xi_1}, \overline{\xi_2}, ...\overline{\xi_n}$. The interconnection between the entropy of physical system $S(\xi_1, \xi_2, ...\xi_n)$ and the statistical weight $W(\xi_1, \xi_2, ...\xi_n)$ of a corresponding state of physical system is defined by the Boltzmann expression $S(\xi_1, \xi_2, ..., \xi_n) = k_B \ln W(\xi_1, \xi_2, ..., \xi_n)$ [27]. The probability of physical system's presence in a particular state, characterized by the parameter $\xi_i$, which is equal to $P_i(\xi_i) = W_i(\xi_i)\big/\sum_{\xi_i} W_i(\xi_i)$. In the equilibrium state, the parameter $\xi_i$ is equal to $\xi_i = \overline{\xi_i} = \xi_{i0}$, and the entropy reaches its maximum value $S_0(\xi_{10}, \xi_{20}, ..., \xi_{n0}) = k_B \ln W_0(\xi_{10}, \xi_{20}, ..., \xi_{n0})$. The thermodynamic fluctuations lead to the deviation of given parameters from their average values. The probability $P$ of the event that the parameter $\xi_i$ has its values in the range $\xi_{i0}+d\xi_i$ is defined by the completed work $dA = -TdS$, which is equal to the change of free energy in the thermodynamic system. At the change from $\xi_{i0}$ to $\xi_i$, the work will be equal to $A = -\int TdS$, or $A = -T(S - S_0) = k_B T \ln(P/P_0)$. The probability of the system's presence in this state is $P = P_0 \exp(-A/k_B T)$. Let us suppose that the deviations $\xi_i - \xi_{i0}$, appearing at the action of random fluctuations, have the alternating signs and small values. Therefore $A$ can be decomposed in the *Taylor* series, which begins with the second order term. Then, we will get $A = b(\xi_i - \xi_{i0})^2 + ...$, where $b$ is the constant. The probability of the system's presence in the state $\xi = \xi_i$ will be

$$P(\xi_i) = P_0 \exp[-b(\xi_i - \xi_{i0})^2/k_B T]. \quad (1)$$

We will obtain the probability magnitude $P_0$ by taking the integer of $P(\xi_i)$ over $\xi_i (-\infty, +\infty)$. These limits can be set, because of the convergence of an integer, which will be equal to the one

$$\int_{-\infty}^{+\infty} P(\xi_i) d\xi_i = P_0 \int_{-\infty}^{+\infty} \exp[-b(\xi_i - \xi_{i0})^2/k_B T] d\xi_i = P_0 \cdot (\pi k_B T/b)^{1/2} = 1. \quad (2)$$

It follows from the above that $P_0 = (b/\pi k_B T)^{1/2}$.

Taking to the consideration that the equilibrium electric field $E$ and magnetic field $H$ are defined by the thermodynamic state of physical body, which is characterized by its temperature, hence in the classic case, the intensities of the electric field $E$ and the magnetic field $H$ have to be among the values of parameters $\xi_i$ [27]. The deviation of the average energy from the equilibrium value may be written as

$$\overline{A} = \frac{1}{2} V_E \varepsilon \varepsilon_0 \overline{(E - E_0)^2} + \frac{1}{2} V_H \mu \mu_0 \overline{(H - H_0)^2}, \quad (3)$$

where $E$ is the intensity of electric field, $H$ is the intensity of magnetic field, $E_0$ and $H_0$ are the equilibrium magnitudes of electric and magnetic fields, $\varepsilon$ is the relative electric permittivity, $\varepsilon_0$ is the free space permittivity, $\mu$ is the relative magnetic permeability, $\mu_0$ is the free space magnetic permeability, $V_E$ is the volume, in which the electric field is concentrated in the electric capacitor with the capacitance $C$, and $V_H$ is the volume, in which the magnetic field is concentrated in the induction $L$. The expressions for the electric and magnetic energies are similar and quadratic by the fields in the both cases. Let us note that there are the resonance phenomena in the $LC$ electrical circuit. The fluctuational oscillations of the electric and magnetic fields have the maximum amplitudes near to the resonance frequency $\omega_0 = (LC)^{-1/2}$ in the $LC$ electrical circuit, while the fluctuational oscillations of the electric and magnetic fields have a wide spectrum similar to the white noise in an electrical capacitor. We will analyze the response of the linear $RC$ electric circuit, and limit our research by the consideration of the case, which is directly related to the experiment [3] and only connected with the electric field $E$ in the capacitor $C$.

The average work $\overline{A}$, which must be done by the system in order to change the intensity of electric field from $E_0$ to $E$, can be represented as the *Gaussian* integral

$$\overline{A} = \frac{1}{2} V_E \varepsilon \varepsilon_0 \overline{(E - E_0)^2} = \frac{1}{2} \int_{-\infty}^{+\infty} V_E \varepsilon \varepsilon_0 (E - E_0)^2 P(E) dE. \quad (4)$$

In this expression, the probability is $P(E) = P_{E0} \exp[-b(E - E_0)^2/k_B T]$, where $P_{E0} = (b/\pi k_B T)^{1/2}$, $b > 0$ and $b = V_E \varepsilon \varepsilon_0 / 2$. The integral, which defines the average work, can be transformed to the expression $\overline{A} = \frac{k_B T}{\pi^{1/2}} I(\varphi)$, where

$$I(\varphi) = \int_{-\infty}^{+\infty} \varphi^2 \exp(-\varphi^2) d\varphi = \pi^{1/2}/2 \quad \text{and}$$

$\varphi = (V_E \varepsilon \varepsilon_0 / 2)^{1/2} (E - E_0)$ (the integral $I(\varphi)$ is shown in [28]). The final result $\overline{A} = k_B T/2$ is equal to the thermal energy, related to the one degree of freedom of the macroscopic body. This is explained by the fact that a part of thermodynamic system, which is connected with electric field $E$, can have the only one predetermined orientation of direction of vector $E$ in every point of space in an electric capacitor. In a plane capacitor, the vector $E$ is directed toward the shortest direction $n_0$ between the two plates, and the problem is effectively reduced to the one dimensional case. In the particular case, it is possible to assume that there is no external electric field, and $E_0 = 0$. At the homogenous temperature, the two orientations of vector $E = \pm E n_0$ have equal probabilities, hence the time dependent fluctuating electric field $E(t)$ is random with alternating plus-minus sign. The average magnitude of electric field is $\overline{E(t)} = \int_{t1}^{t2} E(t) dt / \int_{t1}^{t2} dt = 0$, if the time interval $(t_2 - t_1) >> RC$, where $R$ is some equivalent resistance of the electric circuit with an electric capacitor in Fig. 1.

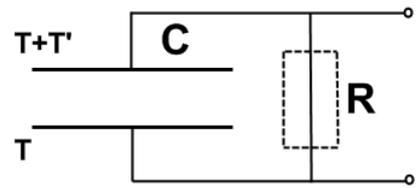

*Fig. 1. Electrical circuit with capacitor C and conditional resistance R; difference of temperatures between plates is $T' \sim T_0 sin(\omega_2 t)$.*

The characteristic time of correlation $\tau = RC$ corresponds to the time of the fluctuation attenuation in the capacitor's electric circuit. Let us note that the resistance $R$ has the temperature $T$, and the fluctuational electric current appears in it in an agreement with the fluctuation-dissipation theorem. The fact that the resistance $R$ doesn't represent a main source of the fluctua-tions of electric potentials difference, but defines the time of relaxation of the fluctuations (see below), is a main distinctive feature of such simple linear scheme. As it was shown above, the average quadratic magnitude of the electric field $\overline{E_t^2} = \int_{t1}^{t2} E(t)E(t)dt / \int_{t1}^{t2} dt$ is not equal to the nil. The full energy of electric field in a capacitor with volume $V_E$ can be expressed by its macroscopic capacitance $C$ as $\frac{1}{2}V_E \varepsilon \varepsilon_0 \overline{E^2} = \frac{1}{2}C\overline{U^2}$, where $U = \int_0^d E(x)dx$ is the electric potentials difference and the capacitance of plate capacitor only depends on the geometric dimensions $C = \varepsilon \varepsilon_0 S_{pl}/d$, where $d$ is the distance between the plates, and $S_{pl}$ is the square of plate of an electric capacitor. Going from this consideration, we come to the expression obtained by *Einstein* [2]

$$\frac{1}{2}C\overline{U^2} = \frac{1}{2}k_B T . \quad (5)$$

This expression doesn't depend on the resistance $R$ in the electric circuit, which connects the plates of an electric capacitor, that is the considered case, when the frequency of thermal fluctuations is significantly bigger than $1/RC$. It is possible to write $\frac{1}{2}q\overline{\overline{U}} = \frac{1}{2}k_B T$, where $q = C\overline{\overline{U}}$ is the average fluctuating electric charge of an electric capacitor, and $\overline{\overline{U}} = \sqrt{\overline{U^2}}$ is the average quadratic value of electric potentials difference. The charge of an electric capacitor at the fixed time moment $q(t)$ includes a discrete number of electrons, while its average quadratic value $\overline{N} = \frac{C\overline{\overline{U}}}{e}$ can be fractional as a result of time averaging ($e$ is the charge of an electron). The stored electric energy is $\frac{1}{2}\overline{N}e\overline{\overline{U}} = \frac{1}{2}k_B T$, and it is proportional to the temperature. If the temperature $T$ and the capacitance $C$ are constants, the average quadratic difference of electric potentials decreases at the increase of a number of fluctuating electrons

$$\overline{\overline{U}} = \left(\frac{kT}{C}\right)^{1/2} = \frac{1}{\overline{N}}\frac{k_B T}{e} \quad (6)$$

and it reaches its maximum at $N = 1$. The electric energy, related to the one fluctuating electron, is $e\overline{\overline{U}} = \frac{k_B T}{\overline{N}}$. The full energy, expressed by the charge and the capacitance, is equal to

$$\frac{1}{2}\frac{\overline{q^2}}{C} = \frac{1}{2}\frac{\overline{N^2}e^2}{C} = \frac{1}{2}k_B T . \quad (7)$$

The same expression between the number of fluctuating electrons and the temperature $\overline{N^2} \propto T$ is true in the metal in the degenerative electron *Fermi*-system at the temperature $T$, which is much less than the temperature of degeneration (see [9], §113). We didn't take to the account the *Fermi* statistics of conduction electrons. The matter is that the electric fields of conduction electrons in the metal are exactly compensated by the charges of metal nucleuses at the distances $r \approx a_0$, where $a_0$ is the distance between the nucleuses; and the condition of electro-neutrality is true in the metal. The translation invariance of metal's crystal grating allows having the free transfer of conduction electrons in the metal. In this case, the *Fermi* statistics has to be taken to the account in view of the big density of states and degeneration of electronic system. The electrons, which are situated near the *Fermi* surface in the momentum space, mainly contribute to the energy and charge transfer. The magnitude of thermoelectric coefficient in metal at low temperature of a few degrees of *Kelvin* is very small $\alpha \propto kT/\varepsilon_F \ll 1$, where $\varepsilon_F$ is the *Fermi* energy [9]. In our case, the electrons, which create both the charge at plates of an electric capacitor and the fluctuating electric field $E$ of an electric capacitor, can be considered as the classic particles, because their surface density at plates of an electric capacitor is not big, hence their *Fermi* energy $\varepsilon_F$ is very small. The characteristic *de Broglie* wavelength of the electrons will be significantly bigger in comparison with the distances between the nucleuses; and the distance between the energy levels is small in comparison with the temperature $\Delta \varepsilon \ll k_B T$. The electric field, created by the electrons, isn't shielded by the charges of nucleuses at the distances between the atoms in the metal, and propagates at the macroscopic distance $d >> a_0$ in the space between the plates of an electric capacitor. In the experiment [3], a total number of the electrons wasn't big indeed (see below). The calculation of the energy of this electric field was completed by the authors above.

Now, let us consider the thermoelectric effect for the fluctuational electrons, appearing in the case, when the plates of an electric capacitor have some different temperatures. In the experiment [3], the plates of an electric capacitor were placed in *Helium II* at the certain equilibrium temperature $T$, and the wave of second sound generated the alternate difference of temperatures $T'(t) = T'_0 exp(i\omega_2 t)$ between the plates ($T'(t) << T$). We will assume that one of the plates of an electric capacitor is at the constant temperature $T$, and the temperature of the second plate is $T + T'(t)$. The change of thermal and electric energies of electrons at the variation of plate's temperature has to be taken to the account in view of the presence of a gradient of temperature $dT'/dr$ between the plates of an electric capacitor. In the linear response theory [27], every electron of $N$ electrons of electric fluctuation takes the additional energy, which is equal to $\frac{1}{2}k_B T'$. This is accompanied by an appearance of the corresponding addition to the electric potentials difference $U'$. Herewith, the change of a full number of the fluctuating electrons $N$ at an action by the small difference of the temperatures $T'$ is also small $N' << N$.

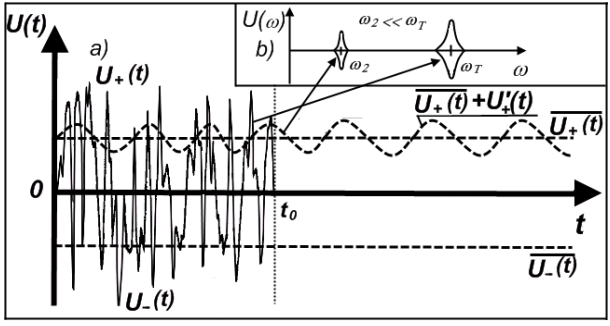

*Fig. 2. Conditional graph of fluctuational dependence of electrical potentials difference U(t) between plates of an electric capacitor on time t at some constant temperature T (a) and on frequency (b), where $\omega_T$ is the Brownian frequency of electric potential difference in electric capacitor and $\omega_2$ is the frequency of second sound in Helium II.*

Let us consider some conditional dependence of electric potentials difference $U(t)$ between the plates of an electric capacitor on the time in Fig. 2a, which has the white noise spectrum. The following expressions to characterize the function $U(t)$ can be written, if the temperature $T$ on the plates of an electric capacitor is same

$$\overline{U(t)} = 0, \quad \overline{U^2(t)} \neq 0, \qquad (8)$$

$$\overline{U_+^2(t)} = \overline{U_-^2(t)} \neq 0, \qquad (9)$$

$$C\overline{U^2(t)} = C\left(\overline{U_+^2(t)} + \overline{U_-^2(t)}\right) = k_B T, \qquad (10)$$

$$C\overline{U_+^2(t)} = C\overline{U_-^2(t)} = k_B T/2, \qquad (11)$$

$$C\overline{U_+^2(t)} = q\overline{\overline{U_+}} = k_B T/2, \qquad (12)$$

where $\overline{\overline{U}} = \left(\overline{U^2}\right)^{1/2}$ and $q = C\overline{\overline{U_+}} = Ne/2$, where $N$ is an average number of fluctuating electrons at the both plates in an electric capacitor.

Let us assume that there are the upper and lower plates in a conditional condensator (Fig. 1). We define the electric potentials difference as $U_+$, when the electrons, creating the electric charge, are at the upper plate in an electric capacitor. We assign the electric potentials difference as $U_-$, when the electrons, creating the electric charge, are at the lower plate in an electric capacitor. Moreover, we assume that the temperature is changing in time and is equal to $T+T'(t)$ at the upper plate of an electric capacitor; and the temperature is constant and is equal to $T$ at the lower plate of an electric capacitor. The oscillating system in [3] can be divided by the two parts: 1) the mechanical part, which is connected with the fluid *Helium II*, and 2) the eletrical part, which is consisted of an electric capacitor in the fluid *Helium II*, and its electrical circuit, including the voltmeter with the input resistance $R$. As it is shown above, the oscillations induced by the wave of second sound in the normal and superfluid helium subsystems are accompanied by the oscillations in the electrical part, which is in the thermodynamic equilibrium with the mechanical subsystem. These oscillations are accompanied by an appearance of the electric potentials difference and currents, flowing by the electric circuit and depending on its resistance $R$. In general case, the correlation function for the voltage $U$ in the linear $RC$ electric circuit depends on the time $t$ as $<U(t_1)U(t_2)> = (k_B T/C)\exp\left[-|t_1-t_2|/RC\right]$. Let us assume that, in the discussed experiments, the time constant of linear electric circuit $RC$ satisfies the condition $|t_1-t_2| << RC$, where $|t_1-t_2| \approx 2\pi/\omega_2$ in [3], and $|t_1-t_2| \approx 2\pi/\Omega$ in [4]. This allows us to simplify the consideration and assume that the exponential term in the correlation function is close to the one (1). Let us think that the heat capacity of metallic plates in an electric capacitor and their near surface areas, where the electric charge is accumulated in an electric capacitor at low temperatures, is small in comparison with the heat capacity of fluid helium, and is not dependent on the thermal state of the system, which is defined by the fluid helium only. We don't count the possible *Kapitsa* temperature jump on the boundary of metal, because the amplitude of the oscillations of temperature $T'$ is significantly smaller than the equilibrium temperature $T$.

In an agreement with the above conditions, it is possible to write an expression for the full thermal and electric energies of an electric capacitor with the plates at different temperatures, taking to the consideration that $T' \ll T$

$$\overline{C(U_+(t)+U'_+(t))^2} + \overline{CU_-^2(t)} = \frac{1}{2}k_B(T+T') + \left(\frac{N}{2}+N'\right)k_B T' + \frac{1}{2}k_B T \quad (13)$$

and for the (+) plate of an electric capacitor at the action of temperature $T'$, it will be equal to

$$\overline{C(U_+(t)+U'_+(t))^2} = \frac{1}{2}k_B(T+T') + \left(\frac{N}{2}+N'\right)k_B T'. \quad (14)$$

The second term in the right part of the equations is connected with the change of a number of electrons at an action of the temperature $T'$. Now, we have to take to the consideration the fact that its frequency is significantly smaller than the frequency of thermal fluctuations, hence it can be considered as an analogue of the quasistationar effect. In this case, a number of fluctuating electrons changes by $N'$ and the energy of each electron changes by $kT'/2$. Let us open the first quadratic term in the left side of equation (14), and considering the difference of times of averaging $\tau \ll \tau'$ for $U_+$ and $U'_+$, let us leave the unaveraging $U'_+(t)$ in the second term. Thus, we will get

$$\overline{CU_+^2} + 2C\overline{\overline{U_+}}U'_+(t) + \overline{CU'^2_+} = \frac{1}{2}k_B(T+T'(t)) + \frac{N}{2}k_B T'(t) + N'(t)k_B T'(t) \quad (15)$$

In the left and right parts of the equation, the second order terms, which have the dashes, are small, and they can be neglected. Then, we will obtain the expression for the first order terms with $U'_+$ and $N'(t)$

$$2C\overline{\overline{U_+}}U'_+(t) = \frac{1}{2}k_B T' + \frac{N}{2}k_B T' = \left(\frac{N+1}{2}\right)k_B T'(t) \quad (16)$$

or

$$NeU'_+(t) = \left(\frac{N+1}{2}\right)k_B T'(t). \quad (17)$$

Let us note that, in the standard theory of measurements [29], the term at the left part of the equation (16), is assumed to be equal to the nill, that is true, if the time of averaging is significantly more than the characteristic time of change of temperature $T'$. However, in [3] and [4], the technique of synchronous detection was used, and theis contribution was registered at second sound frequency of $\omega_2$, which is characteristic for $T'$, that is why it was not averaged and is different from the nill. Going from this, we will obtain the final result

$$\alpha = \frac{U'_+(t)}{T'(t)} = -\frac{(N+1)}{2N}\frac{k_B}{|e|} = -\frac{k_B}{2|e|} - 0\left(\frac{1}{2N}\right). \quad (18)$$

The equation (18) corresponds to the thermoelectric coefficient $\alpha$ for the electrons, which take part in the fluctuation process at presence of the changing temperatures difference $T'(t)$ between the plates in an electric capacitor.

This result quantitatively corresponds to the magnitude of the thermoelectric coefficient $\alpha$, obtained in the experiment in [3]. It also follows from the expression (18) that the negative electrons charge is accumulated at the heated plate in an electric capacitor at increase of temperature $(T'>0)$. In the experiment [3], the plate with the high temperature was negatively charged in an electric capacitor.

Thus, the creation of small enough oscillations of temperatures difference $T'_{\omega 2} \propto T'_0 \sin(\omega_2 t)$ by the wave of second sound in the fluid $^2He$ has to be accompanied by an appearance of the additional in-phase electric potentials difference $U'_{\omega 2} = \alpha T'_0 \sin(\omega_2 t)$, where $\alpha = -k_B/2|e| \approx 4.3126 \cdot 10^{-5}$ V/K. The equilibrium temperature $T$ is not included in equation (18), hence the results don't depend on the magnitude of equilibrium temperature $T$, as it was observed in [3]. In the researched case [3], the alternate electric field reached the magnitude $E'_0 \sim (10^{-4} \div 10^{-6})$ V/m in an electric capacitor. The helium atoms didn't contribute a notable change in the obtained result, because the polarizational effects for *Helium II* atoms are proportional to the square of intensity of electric field as in the case of all the inert gases with the filled electron shells.

### Oscillations of difference of electric potentials in electric capacitor in mechanical torsional resonator

Let us analyze the results of experiment [4], in which the oscillations of electric potentials difference $U'(t)$ at the electrodes of the cylindrical electric capaci-tor (Fig. 3), connected with the mechanical torsional resonator, were observed in the range of temperatures from $1.4\ K$ to $T_\lambda = 2.172\ K$.

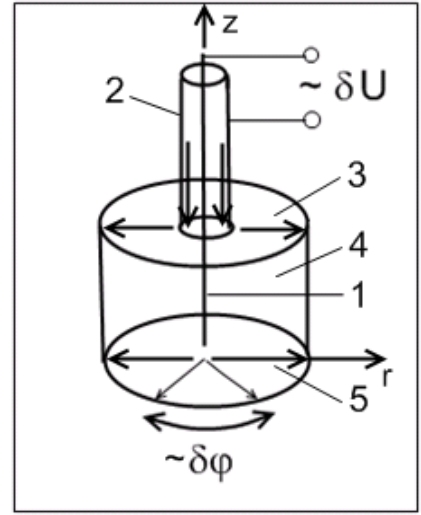

*Fig. 3. Scheme of mechanical torsional resonator with fixed metallic electrode (1), and oscillating metallic camera (2, 3, 4, 5) (after [3]). Arrows denote direction of flow of superfluid component of helium at moment, when velocity of rotation of resonator is maximal.*

In Fig. 3, the first plate of an electric capacitor includes the fixed electrode *(1)*; the second plate of an electric capacitor consists of the cylindrical surfaces *(2)* and *(4)* with the covers *(3)* and *(5)*, involved in the rotational movement. The maximum amplitude of electric potentials difference was $U' \approx 1.5 \cdot 10^{-7}$ V at the temperature $T \approx 2\ K$ [4]. The *Helium II* as the liquid or as the saturated or not saturated film, covering the inside surface of a resonator, was placed in a resonator. In an agreement with the equation (18), the oscillations of the electric potentials difference at such a scale can be originated by the oscillations of the temperatures difference $T' \sim 10^{-3} K$ in a given capacitor. Considering the dependence of relative concentration of superfluid component on the temperature, we can find that, at the temperature $T = 2K$, the magnitude is $\rho_s/\rho \approx 0.3$, where $\rho_s$ is the density of superfluid component, and $\rho$ is the full density. In agreement with the data in [30], at the above temperature $T$, the change of density of the superfluid component $\rho'_s$, appearing at the action by the oscillations of given temperatures difference $T'$, doesn't exceed the magnitude $\rho'_s/\rho \approx 4 \cdot 10^{-3}$.

Let us consider the case, when there is a saturated film of superfluid helium inside the resonator. As it is known [30], the thickness of saturated superfluid helium film is $\sim (25-30)\ nm$, and it contains the *100-120* layers of the helium atoms approximately, and it covers all the inside surface of a resonator. The thickness of film slightly decreases in the upper part of a resonator, but this effect has no principal influence in this case, hence we don't consider it. It is visible that the change of density $\rho'$ is commensurable with the mass of approx-imately *0.4–0.5* single layer of helium at the above described oscillations of temperature. The flow of small amount of the superfluid component from one part of a resonator to another part of a resonator will result in the change of distribution of temperatures in a resonator at the adiabatic conditions due to the above considered mechanism, resulting in the observed effect [4].

Let us define the physical mechanism and forces, which have the influences on the dynamics of the normal and superfluid components of *Helium II*, and their transport during the rotation of a resonator in the considered research task. In the paper [4], the resonator conducted the rotational oscillations $\varphi(t)=\varphi_0 sin(\Omega t)$, where $\varphi_0$ is the amplitude of angular oscillations, expressed in the radians, $\Omega$ is the angular frequency. The velocity of rotation of resonator's walls is $\upsilon_\varphi(r)=rd\varphi(t)/dt=r\varphi_0\Omega cos(\Omega t)$, where $r$ is the radius of given point of rotating surface. At the surface (4) in Fig. 3, the velocity of rotation had a maximum value [4]. It is possible to assume that, in view of the big viscosity and small thickness of the superfluid film, the normal helium component is fully dragged by the rotating resonator's walls, taking part in the rotational movement with the velocity $\upsilon_{n,\varphi}(r,t)$, which is equal to the velocity of walls rotation $\upsilon_\varphi(r)$. The centrifugal force $F_c=\rho_n(\upsilon_{n,\varphi})^2/r$, acting on the normal component, can be regarded as small enough in the considered case, assuming that the radial velocity of normal component is $\upsilon_{n,r}\approx 0$. The superfluid component is not involved in the rotational movement by the resonator's walls with the velocity smaller than the critical velocity of vortices generation, that is $\upsilon_{s,\varphi}=0$. Let us note that, during the oscillations of a resonator, the superfluid component can make the oscillating movements with some radial velocity $\upsilon_{s,r}\neq 0$, and it can flow from the regions with the small radius $r$, that is from the surface (2) and from the central regions of covers (3) and (5) to the side surface (4), which has the biggest radius $R$, and then flowing backward. Let us clarify the cause of oscillating movement origination. In some sense, the oscillations of such a type are similar to the oscillations between the two connected vessels, when the *Helium II* levels inside the vessels are close to the equilibrium [31], and the vessels are connected by the thin constriction or when the interflow of the *Helium II* between the vessels is done by the means of the superfluid helium film at presence of the outer pressure.

The system of hydrodynamic equations for *Helium II* was firstly proposed by *Landau*, and then developed by *Khalatnikov* (see [32, 33, 34]). Let us write the equation of movement for the superfluid component at coordinate $r$

$$\rho_s\frac{\partial \upsilon_s}{\partial t}+\rho_s\nabla_r\left(\mu+\frac{\upsilon_s^2}{2}-\frac{\rho_n(\upsilon_{n,\varphi}-\upsilon_s)^2}{2\rho}\right)=0, (19)$$

where $\mu$ is the chemical potential. Let us assume that the rotational oscillations are generated with the small angular amplitude and with the small linear velocity $\upsilon_\varphi$, which is less then the critical velocity of vortex generation. In the initial moment, the superfluid component will be in the state of quiescence with the velocity $\upsilon_s=0$, and by selecting the phase of oscillations, when $\nabla\mu=0$, the equation (19) can be written as

$$\rho_s\frac{\partial \upsilon_{s,r}}{\partial t}=\rho_s\nabla_r\left(\frac{\rho_n}{2\rho}\upsilon_{n,\varphi}^2\right). \; \upsilon_n^2=[r\varphi_0\Omega cos(\Omega t)]^2 \; (20).$$

Let us consider the above written dependence of $\upsilon_{n,\varphi}(\Omega)$ and derive the final expression for the force, acting on the superfluid component as

$$\rho_s\frac{\partial \upsilon_{s,r}}{\partial t}=\frac{\rho_s\rho_n}{\rho}r\varphi_0^2\Omega^2\left(1+\cos(2\Omega t)\right) \qquad (21)$$

The right parts in the equations (20, 21) represent the *Bernoulli* force, which acts on the superfluid component of *Helium II*, transporting it to the part of a resonator, where the linear velocity $\upsilon_{n,\varphi}$ has its maximum value. It is well known fact that the *Bernoulli* force can not originate in the experiments to research the *Venturi* effect in *Helium II* in the pipe with constriction [35-37]. In an agreement with [37], in this case, the condition $rot\upsilon_s=0$, where the direction $\upsilon_s$ coincides with the direction of movement of the normal component $\upsilon_n$ of *Helium II* in the pipe, has to be true for the superfluid component of *Helium II*. This results in the equality of *Helium II* levels, measured in the region of wide pipe and in the region of the constriction of wide pipe. In the researched case, the appearing superfluid component velocity $\upsilon_{s,r}$ is directed along axe $r$ and is orthogonal to the normal component velocity $\upsilon_{n,\varphi}$ in the equation (16). At these conditions, the *Bernoulli* effect has to be originated, because the expression $rot\upsilon_s=0$ is true in view of the existing orthogonality between the velocity vector and the direction of circular contour bypassing. The interflow of the superfluid component, which doesn't transfer the entropy, results in an appearance of the difference of temperatures $T'$ between the resonator's regions with the small and big coordinates $r$. In the turning points of a resonator, the velocity of normal component is $\upsilon_{n,\varphi}=0$, and the *Bernoulli* force is equal to the nil. The thermomechanical effect, which creates a contrary stream of the superfluid *Helium II* and results in an origination of the oscillation process, plays a main role at the turning points of a resonator. The phase of oscillations of *Helium II* can be shifted in relation to the resonator's oscillations, because of the inertia of the flow of mass, however this effect can be disregarded in view of the small mass transfer. These forced oscillations of the temperatures difference and the flow of superfluid component at the axe $r$ can transform to the fading oscillations, continuing for some time after the moment of resonator's rotation stopping, as observed in [4].

Going from the calculations, the contribution by a cylindrical capacitor with the small radius with the electrodes (1) and (2), and the covers of a big volume resonator with the electrodes (1) and (3) and the electrodes (1) and (5) to the full value of capacitance of a sectional capacitor exceeds the capacitance of a capacitor with the electrodes (1) and (4) in more than *5* times. Therefore, namely the plates (2), (3) and (5) will create the main electric response of a sectional capacitor. In the connection with the interflow of superfluid component from these surfaces to the surface (4) during the oscillations, all the plates will experience the increases of temperatures periodically. Therefore, the plates will charge negatively in an agreement with the equation (18), but the central electrode (1) will charge positively, because it will have the lower temperature, comparing to the other plates. To account for the fluctuational contribution, let us represent the variables as $T = T_0 + T'$ and $\upsilon_n = \upsilon_{n0} + \upsilon'$, where the part with the index 0 corresponds to the equilibrium value, but the second part corresponds to the value, which depends on the time and originates as a result of the oscillations of a resonator. Let us evaluate the amplitudes of temperature difference $T'$, appearing between the central and distant regions of a resonator.

Let us substitute $d\mu = -\sigma dT + dP/\rho$, where $\sigma = S/\rho$, in the equation (19)

$$\rho_s \frac{d\boldsymbol{\upsilon}_s}{dt} + \frac{\rho_s}{\rho}\nabla P + \frac{\rho_s}{2}\nabla \boldsymbol{\upsilon}_s^2 - \rho_s \sigma \nabla (T_0 + T') - \frac{\rho_s \rho_n}{2\rho}\nabla (\upsilon_{n,0} + \upsilon_{n,\varphi} - \boldsymbol{\upsilon}_s)^2 = 0 \quad (22)$$

Let us assume that the gradient of external pressure $\nabla P \approx 0$, and also the following relation is true $\upsilon_s / \upsilon_n \ll 1$ and $\nabla \upsilon_s^2 \approx 0$ in an agreement with the conditions of experiment. In the extremum points $d\boldsymbol{\upsilon}_s/dt = 0$, and we can derive the inter-coupling expression between the temperature and the velocity of normal component of *Helium II* by integrating the two last terms in the equation (22) over the $r$ (see the similar solution for the thin capillary in [32] §140). From the equation (22) we will obtain

$$\frac{\rho_s \rho_n}{2\rho}(\overline{\overline{\upsilon_{n0}}} + \upsilon_{n,\varphi})^2 = -\rho_s(T_0 + T')\sigma. \qquad (23)$$

In this case $\upsilon_{n,\varphi} \ll \overline{\overline{\upsilon_{n0}}}$ where $\overline{\overline{\upsilon_{n0}}}$ is the average thermal velocity of the helium atoms, $\upsilon_{n,\varphi}$ is the velocity of the normal component on the surface of rotation. For the numerical evaluation of the effect's magnitude, let us write the expression (23) for a capacitor with small diameter (the surface number (2), Fig. 1) and a capacitor with big diameter (the surface number (4), Fig. 1), taking to the consideration the volumes $V_1$ and $V_2$ of films of fluid *Helium II*, situated in every capacitor. Let us assume that, in the adiabatic case in the corresponding phase of wave process, the superfluid component of *Helium II*, which flows from the small capacitor, results in an increase of temperature at the surface (2) and in a small change in the kinetic term at left part of the equation (23), because of the small magnitude of surface radius and small velocity of its rotation. This superfluid helium, inflowing into a big capacitor at the surface (4), leads to the small decrease of temperature in view of its small volume in comparison with the volume $V_2$, situated inside it, but the term, connected with the kinetic energy, is significantly increased, because of big velocity of rotation of the surface. Let us assume that the linear velocity of rotation doesn't increase above the critical value, hence the superfluid component is not trapped by the oscillating resonator and $\upsilon_{s,\varphi} = 0$. Let us sum up the equations (23) for the both considered capacitors, leaving the main terms, which have an influence on the change of temperature only. Then, as in the case of the equation (16), we will obtain

$$V_2 \frac{\rho_s \rho_n}{2\rho}(\overline{\upsilon_{n0}^2} + 2\overline{\overline{\upsilon_{n0}}}\upsilon_{n,\varphi} + \upsilon_{n,\varphi}^2) = -V_1 \rho_s (T_0 + T')\sigma, (24)$$

where $\upsilon_{n,\varphi}$ is the velocity of the normal component of *Helium II* on the rotation surface (4) (Fig.1).

Let us believe that the value $\upsilon_{n\varphi}^2$ is quadratically small, and it can be disregarded. Equalizing the linear $T'$ and $\upsilon_{n\varphi}$ terms, we will find the dependence of amplitude of oscillations of the temperature $T'$ in a small capacitor on the velocity of rotation of surface $\upsilon_{n\varphi}$ in a big capacitor (4) as

$$T' = -V_2 \rho_n \overline{\overline{\upsilon_{n0}}} \upsilon_{n,\varphi} / V_1 \rho \sigma. \qquad (25)$$

Let us conduct the qualitative evaluation of amplitude of oscillations of temperature, for example, at the temperature $T = 2K$. In [4], the maximum linear velocity of rotation was $\upsilon_{n,\varphi} = 7 \cdot 10^{-4} m/sec$. At the given temperature the thermal velocity for the helium atoms is $\overline{\overline{\upsilon_{n0}}} \approx 100 m/sec$, the specific entropy of *Helium II* is $\sigma \approx 940 \; J/kg \cdot K$, $\rho_n/\rho = 0.7$, and the relation between the volumes of *Helium II* films on the surfaces (2) and (4) is $V_2/V_1 \approx 16.7$. Then, going from the equation (25), we will obtain $T' \approx 1.3 \cdot 10^{-3} K$. The oscillations of difference of electric potentials will be mainly defined by the change of temperature in a capacitor with small radius (electrodes (1) and (2)), and in an agreement with the equations (18) and (25), we will obtain

$$U'_{2\Omega} = \frac{k_B}{2|e|} V_2 \rho_n \overline{\overline{\upsilon_{n0}}} \upsilon_{n,\varphi} / V_1 \rho \sigma. \qquad (26)$$

This result is in a good qualitative agreement with the data, obtained in [4]. From the equation (21), it follows that $\upsilon_{n,\varphi} \sim r\varphi_0 \Omega \cos(\Omega t)(1 - \sin^2(\Omega t))$, and the maximum value of the electric potentials difference must be observed at the maximum deflection angle of a resonator, but not at the moment, when the maximum value of acceleration is reached, as confirmed in [4].

Thus, the rotation of an oscillator in [4] results in both an increase of the normal component with subsequent interflow of superfluid component of *Helium II* at the *Bernoulli* force action on the resonator's wall (4) with the biggest radius of rotation as well as an appearance of the biggest difference of temperatures between the plates of a capacitor with the small radius of rotation (1-2). In the experiments [3] and [4], the physical foundations of observed processes are based on the thermoelectric effect, appearing for the *Einstein's* fluctuating electrons at the presence of the temperatures difference between the plates of a capacitor.

## Discussion on theoretical research results

Going from the research by *Einstein* [1], the authors proposed the theory of thermoelectric effect for the fluctuating electrons, allowing to explain the results of experiments in [3] and [4], which are connected with the thermal effects, accompanied by an appearance of the small electric fields in an electric capacitor in *Helium II*. The thermoelectric coefficient α is equal to the value, which corresponds to the value in the known classic theory by *Drude* [38], and approximately in $10^4$ times bigger, than it can be in the metals at such low temperatures, where it is reduced on $k_B T/\varepsilon_F$. Using the expression (7), it is possible to evaluate a total number of fluctuating electrons in the experiment [3], where the capacitance of a capacitor was $C \approx 5 \cdot 10^{-13} F$ [39]. We obtain that the total number of fluctuating electrons is $\bar{\bar{N}} \approx 23$ at temperature $T = 2K$. At $T = 1,4K$, the total number of fluctuating electrons is $\bar{\bar{N}} \approx 19$. In the researched case, the average number of fluctuating electrons is small, and they can be considered as an ensemble of classic particles, without taking to the account the *Fermi* statistics. The change of temperature fields in *Helium II* is directly connected with the transfer of the superfluid and normal components of *Helium II*, and observed at the propagation of waves of second and third sounds or at the interflow of superfluid films [31, 33].

The dielectric constant $\varepsilon$ of the fluid *Helium II* depends on the temperature, and it can have an influence on the capacitance magnitude of a capacitor. At the change of temperature from $T_\lambda$ to $1K$, the relative value of change of density is approximately $\Delta\rho/\rho \approx 10^{-3}$, therefore the change of $\Delta\varepsilon/\varepsilon$ has the same value. The correction of coefficient $\alpha$ in this mechanism is near $10^{-8} V/K$, that is why it was not taken to the consideration in this case. Undoubtedly, it is necessary to take to the account this correction in the case of the experiments with the big magnitudes of electric fields.

In our opinion, the maximum $U'$ at the temperature of around $2K$ (Fig. 3) in [4] is connected with the maximum of amplitude of oscillations of heat flux $W' = \rho C T' u_2$ in the case of second sound propagation, where $C$ is the heat-capacity of *Helium II*. During the propagation of superfluid film, the maximum $W'$ is in the same temperatures range, though the velocity of superfluid film propagation is approximately one order of magnitude less than the velocity of second sound propagation.

We don't provide the exact numerical analysis of frequency spectrum in [3], because it strongly depends on a number of conditions of the experiment, which are not described in [3].

Let us draw attention to the fact that the clear difference of amplitudes $U'$ at the rotation of a resonator in different directions is visible on the dependence of the amplitude of oscillations $U'$ on the time $t$, described by the equation (21) (see Fig. 2) in [4]. This difference of amplitudes $U'$ can be explained by the varying velocity of normal component of helium $\upsilon_{n,\varphi}$, appearing in view of the non-symmetric polishing of resonator's plates, but not because of the presence of some circulating superfluid *Helium II* stream, as it was assumed by the researchers [4]. In the case of circulation, the time semi-periods of oscillations, connected with the rotation of a resonator in the different directions, must not be equal, however they are equal precisely in [4].

Let us pay attention to the fact that the generation of vortices at the velocities above the critical velocity results in a suppression of the electric effect [4], because the magnitude of flowing superfluid *Helium II* stream, the oscillations of difference of concentrations of *Helium II* components, the thermal and electric effects are significantly limited, because both the superfluid component of *Helium II* as well as the normal component of *Helium II* take part in the rotational movement during the process of the vortices generation, which is characterized by the average value expression: $<\mathrm{rot}\,\upsilon_{s,\varphi}> \neq 0$.

In this research, we don't consider the effect [3], connected with the generation of oscillations of second sound, when the alternate difference of electric potentials $U'$ with the magnitude in $10^8 - 10^{10}$ times bigger, than the magnitude in this research, was applied to the additional capacitor inside the second sound resonator, reaching the electric field magnitude $E \approx 1,67 \cdot 10^4 V/m$. In this case, the electric field can not be considered as small enough, and its influence on the properties of superfluid *Helium II* have to be taken to the account. The different research approach has to be used to interpret this effect, which can be discussed in our next research paper.

## Conclusion

The theory of thermoelectric effect with the fluctuational electrons at the plates of an electric capacitor in the superfluid *Helium II* with the oscillations of the second sound wave is proposed. The results of theoretical calculations are in good agreement with the experimental data, obtained at the research of electric signals in both the second sound resonator and the rotational torsional mechanical resonator in [3, 4]. The similar described thermoelectric effect can be realized in the capacitors at the origination of the temperatures difference between the plates of a capacitor at various temperatures. The thermoelectric effect can be used in the numerous measurement systems in view of a big magnitude of thermoelectric coefficient. In the authors' opinion, the capacitor represents an effective thermoelectric transducer of a new type.

Authors express thanks to A. S. Rybalko for the thoughtful discussion and elaboration on the details of experiments in [3, 4].

This innovative research paper was published in the *Problems of Atomic Science and Technology* (*VANT*) in *2014* in [40].

*E-mails: ledenyov@kipt.kharkov.ua,
ledenyov@univer.kharkov.ua .

———————

1. A. Einstein, Theorie der Brownschen Bewegung. *Ann., Phys.*, **19**, p. 371 (1906).
2. A. Einstein, *Ann. Phys.*, v. **22**, p. 569 (1907).
3. A. S. Rybalko, *Low. Temp. Phys.*, v. **30**, iss. 12, p. 994 (2004).
4. A. S. Rybalko, S. P. Rubets, *Low. Temp. Phys.*, v. **31**, iss. 7, p. 623 (2005).
5. D. L. de Haas-Lorentz, *Die Brownsche Bewegung und einige werwandte Erscheinungen*, Verl. Fr. Vieweg., *Die Wissenschaft, B.*, v. **52** (1913).
6. H. Nyquist, *Phys. Rev.*, v. **32**, p. 110 (1928).
7. H. B. Callen, T. A. Welton, *Phys. Rev.*, v. **83**, no.1, p. 34 (1951).
8. V. L. Ginzburg, *Usp. Fiz. Nauk*, v. **46**, iss.3, p. 348 (1952) (in Russian).
9. L. D. Landau, E. M. Lifshitz, *Statistical Physics*, Part 1, *Pergamon Press* (1980).
10. A. I. Akhiezer, *Electrodynamics of plazma*, Moscow, *Nauka* (in Russian) (1974).
11. E. M. Lifshitz, L. P. Pitaevskii, *Statistical Physics*, Part 2, *Pergamon Press* (1981).
12. Yu. S. Barash, V. L. Ginzburg, *Usp. Fiz. Nauk*, v. **116**, iss.1, p. 5 (1975) (in Russian).
13. Yu. S. Barash, *Van der Vaals force*, Moscow, *Nauka* (1988) (in Russian).
14. V. D. Khodusov, Vestnik Kharkovskogo Universiteta, physical series "Nucleus, particles, fields", № **642**, iss. 3, p. 12 (2004).
15. A. M. Kosevich, *Low Temp. Phys.*, v. **31**, iss. 1, p. 37 (2005).
16. L. A. Melnikovsky, arXiv:cond-mat/0505102v3 (2007); *J. Low Temp. Phys.*, v. **148**, p. 559 (2007).
17. A. M. Kosevich, *Low Temp. Phys.*, v. **31**, iss. 10, 1100(839) (2005).
18. V. D. Natsik, *Low Temp. Phys.*, v. **31**, iss. 10, 1201(915) (2005).
19. D. M. Litvinenko, V. D. Khodusov, Vestnik Kharkovskogo Universiteta, physical series "Nucleus, particles, fields", № **721**, iss. 1, p. 31 (2006).
20. E. A. Pashitskii, S. M. Ryabchenko, *Low. Temp. Phys.*, v. **33**, iss. 1, 12(8) (2007).
21. V. M. Loktev and M. D. Tomchenko, *Low Temp. Phys.*, v. **34**, iss. 4/5, 337(262) (2008).
22. E. A. Pashitskii, O. M. Tkachenko, K. V. Grygoryshyn, B. I. Lev, *Ukr. Fiz. Zhurn.*, v. **54**, №1-2, p. 93 (2009) (in Ukrainian).
23. S. I. Shevchenko, A. S. Rukin, *JEPT Letters*, v. **90**, № 1, 46(42) (2009).
24. E. D. Gutliansky, *Low Temp. Phys.*, v. **35**, iss. 10, 956(748) (2009).
25. S. I. Shevchenko, A. S. Rukin, *Low Temp. Phys.*, v. **36**, iss. 2, 186(146) (2010).
26. E. A. Pashitskii, A. A. Gurin, *ZhETF*, v. **138**, iss. 6(12), p. 1103 (2010) (Russian).
27. R. Kubo, *Statistical mechanics*, North-Holland Publ. Com. (1965).
28. I. S. Gradshtein, I. M. Ryzhik, *Tables of integrals, series and products*, 5 ed., AP, p.382, №3.461(2) (1996).
29. V. B. Braginsky, F. Ya. Khalili, *Quantum measurement*, Cambridge University Press (1992).
30. B. N. Esel'son, V.N. Grigoryev, E.Y. Rudavskiy, *Properties of liquid and solid helium*, Moscow (1978) (in Russian).
31. K. R. Atkins, *Liquid helium*, Cambridge University Press (1959).
32. L. D. Landau, E. M. Lifshitz, *Fluid Mechanics*, 7th ed., *Pergamon Press* (1987).
33. I. M. Khalatnikov, *An Introduction to the Theory of Superfluidity*, Westview Press (2000).
34. D. R. Tilley, J. Tilley *Superfluidity and superconductivity*, Department of Physics, University of Essex, *Van Nostrand Reinhold Company* (1974).
35. R. Meservey, *Phys. Fluids*, v. **8**, p. 1209 (1965).
36. W. M. Van Alphen, R. de Bruyn Ouboter et al., *Physica*, v. **39**, p. 109 (1968).
37. S. J. Patterman, *Superfluid Hydrodynamics*, North-Holland Publ. Com. (1974).
38. N. W Ashkroft, N.D. Mermin , *Solid State Physics*, *Saunders College Publishing*, Philadelphia, USA (1976).
39. A. S. Rybalko, Private communications, Kharkov, Ukraine, 2012.
40. Dimitri O. Ledenyov O, Viktor O. Ledenyov, Oleg P. Ledenyov Electrical effects in superfluid helium. 1. Thermoelectric effect in Einstein's capacitor, *Problems of Atomic Science and Technology* (*VANT*), Series «Vacuum, pure materials, superconductors», no 1(89), pp. 170 - 179, ISSN 1562-6016, (2014) http://vant.kipt.kharkov.ua/ARTICLE/VANT_2014_1/article_2014_1_170.pdf .